\documentclass{article}
\usepackage{spconf,amsmath,graphicx,hyperref}
\usepackage{multirow}
\usepackage{arydshln}
\usepackage{bm}
\usepackage{url}
\usepackage{amsmath}
\usepackage{amssymb}
\newcommand{\bmit}[1]{{\mbox{\boldmath $#1$}}}

\title{Investigation of generalization to unseen audio generation models in audio--text semantic-alignment score prediction}
\name{Author(s) Name(s)\thanks{Thanks to XYZ agency for funding.}}
\address{Author Affiliation(s)}

\name{\shortstack{Yuki Okamoto$^1$, Noriyuki Tonami$^{2}$, Takao Kawamura$^{2}$, Keisuke Imoto$^{3}$,\\Ryotaro Nagase$^{4}$, Riki Takizawa$^{5}$, Minoru Kishi$^{6}$, and Shinnosuke Takamichi$^{6,1}$}}
\address{$^{1}$The University of Tokyo, Japan, $^{2}$NEC Corporation, Japan, $^{3}$Kyoto University, Japan,\\$^{4}$Ritsumeikan University, Japan, $^{5}$Kyoto Sangyo University, Japan, $^{6}$Keio University, Japan,\\ }
\begin{document}
\ninept
\maketitle
\begin{abstract}
Text-to-audio generation (TTA) requires reliable evaluation of semantic alignment between generated audio and input text.
Existing learning-based alignment-score prediction models have been evaluated only on audio generated by models included in the training data, leaving their generalization to unseen generation models unexplored.
In this paper, we investigate the generalization of audio--text semantic-alignment score prediction models to unseen TTA models and introduce adversarial training using a generation-model classifier with a gradient reversal layer (GRL).
Experimental results show that pretrained audio encoders capture generation-model-specific characteristics and that alignment-score prediction performance degrades on audio generated by unseen TTA models.
Applying the GRL improves prediction performance in some settings, although its effectiveness in suppressing generation-model-specific information is limited.
\end{abstract}
\section{Introduction}
Text-to-audio generation (TTA), which generates environmental sounds from textual descriptions, has attracted considerable research attention \cite{Liu_TASLP_2024}.
High-quality audio generation systems have great potential for many applications, such as supporting sound production for movies and video games \cite{Lloyd_SIGG_2011}, and data augmentation for environmental sound analysis \cite{Kong_ICASSP_2019}.

In evaluating TTA models, semantic alignment between the input text and generated audio is an important metric \cite{Lee_NeurIPS_01}.
Several objective metrics and learning-based methods have been proposed to automatically evaluate audio--text semantic alignment \cite{interspeech2025_kanamori_01}.
However, existing alignment-score prediction models have been evaluated mainly on audio generated by the same audio generation models included in the training data.
As TTA models become increasingly diverse, alignment-score prediction models need to generalize to audio generated by unseen audio generation models.

In this study, we first investigate whether generated audio exhibits generation-model-specific characteristics.
We then analyze the robustness of alignment-score prediction models when the audio used for training and evaluation is generated by different generation models. 
Finally, we investigate the effectiveness of adversarial training using a generation-model classifier and a gradient reversal layer for improving generalization to unseen models.

\section{Related work}
\textbf{Audio--text semantic-alignment score prediction.}
Several objective metrics have been proposed for evaluating audio--text semantic alignment.
CLAPScore measures semantic alignment using the cosine similarity between audio and text embeddings obtained from a pretrained contrastive language–audio pretraining (CLAP) model \cite{Wu_ICASSP_2023}.
However, CLAPScore has been reported to exhibit only a weak correlation with human subjective evaluations.
To achieve a higher correlation with subjective evaluations, Kanamori et al.  \cite{interspeech2025_kanamori_01} proposed a supervised learning method that predicts audio--text alignment scores from audio and text using subjective scores as supervision.
An international competition on audio--text semantic-alignment score prediction has also been organized \cite{Okamoto_ICASSP_2026}, resulting in several publicly available prediction models.
However, these prediction models have not been evaluated on audio generated by unseen audio generation models.

\begin{figure*}[t!]
    \centering
    \includegraphics[width=0.85\linewidth]{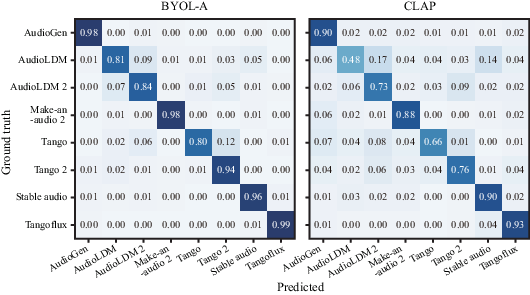}
    \caption{Confusion matrix of TTA models classification}
    \label{fig:confusion_matrix}
\end{figure*}
\textbf{MOS prediction for unseen speech synthesis models.}
Generalization to unseen generation models has been investigated in automatic mean opinion score (MOS) prediction for speech.
Previous studies \cite{Cooper_ICASSP_2022} have reported that MOS prediction performance degrades when evaluating synthetic speech generated by models not included in the training data.
Huang et al. \cite{Huang_ICME_2026} introduced a gradient reversal layer (GRL) \cite{Ganin_ICML_2015} into an automatic MOS prediction model to suppress generation-model-specific information and improve robustness to unseen generation models.
Motivated by this approach, we investigate the effects of adversarial training with a GRL on the generalization of audio--text semantic-alignment score prediction to unseen audio generation models.

\section{Analysis of generated audio samples}
\label{sec:pre_exp}
Audio generated by text-to-audio (TTA) models may contain model-specific characteristics.
If such characteristics exist, an audio--text semantic-alignment score prediction model may rely on them when predicting alignment scores. As a result, its prediction performance may degrade on audio generated by unseen TTA models.
Therefore, we first investigate whether synthesized audio contains characteristics that allow the generating TTA model to be identified.

To investigate the existence of model-specific characteristics, we trained a classifier to identify the TTA model that generated each audio sample.
The classifier consists of a pretrained audio encoder and a two-layer multilayer perceptron (MLP).
During training, only the MLP parameters were updated, while the audio encoder was frozen.

We used the XACLE Challenge 2026 dataset, which contains audio generated by eight different TTA models.
We employed BYOL-A and CLAP as audio encoders.
The classifier was trained using the Adam optimizer with a batch size of 256 and a learning rate of 0.001.

Figure \ref{fig:confusion_matrix} shows the classification results for audio generated by the eight TTA models.
The results demonstrate that the generating TTA models can be identified with high accuracy using only the synthesized audio, indicating the presence of model-specific characteristics.

\begin{table*}[t!]
\caption{Results of semantic-alignment score prediction.}
\label{table:result_baseline}
\centering
\setlength{\tabcolsep}{8pt}
\resizebox{\linewidth}{!}{%
\footnotesize
\begin{tabular}{@{}c|l||rrrr|rrrr@{}}
    \hline
      \multirow{3}{*}{Training data} &\multirow{3}{*}{Method} & \multicolumn{8}{c}{Test data}\\
      &  & \multicolumn{4}{c}{A} & \multicolumn{4}{c}{B} \\
      \cline{3-10}
       &  & SRCC $\uparrow$ & KTAU $\uparrow$ & LCC $\uparrow$ & MSE $\downarrow$ & SRCC $\uparrow$ & KTAU $\uparrow$ &LCC $\uparrow$ & MSE $\downarrow$\\
    \hline
    \multirow{4}{*}{A} & Baseline & $0.339$ & $0.232$ & $0.340$ & $5.928$ & $0.144$ & $0.097$ & $0.146$ & $5.870$\\
    \cdashline{2-10}
      & Ours &&&&&&&&\\
       & ~~~All & $\mathbf{0.413}$ & $\mathbf{0.284}$ & $\mathbf{0.403}$ & $\mathbf{5.141}$ & $\mathbf{0.161}$ & $\mathbf{0.107}$ & $0.147$ & $5.681$\\ 
       & ~~~Only audio encoder & $0.262$ & $0.178$ & $0.269$ & $5.722$ & $0.160$ & $0.105$	& $\mathbf{0.156}$ & $\mathbf{5.415}$\\
     \hline
    \multirow{4}{*}{B} & Baseline & $0.210$ & $0.141$ & $0.210$ & $\mathbf{5.747}$ & $0.393$ & $0.269$ & $0.376$ & $4.524$\\
    \cdashline{2-10}
      & Ours &&&&&&&&\\
      & ~~~All& $\mathbf{0.241}$ & $\mathbf{0.161}$ & $\mathbf{0.234}$ & $5.756$ & $0.398$ & $0.272$ & $0.375$ & $4.513$\\ 
      & ~~~Only audio encoder & $0.204$ & $0.138$ & $0.207$ & $5.972$ & $\mathbf{0.412}$ & $\mathbf{0.283}$ & $\mathbf{0.396}$ & $\mathbf{4.481}$\\
    \hline
\end{tabular}
}
\end{table*}
\section{Generalization analysis of semantic-alignment score prediction models}
\subsection{Baseline architecture}
We used the official baseline model of the XACLE Challenge 2026 \cite{Okamoto_ICASSP_2026} for audio--text semantic-alignment score prediction.
This model consists of an audio encoder (pre-trained BYOL-A \cite{Niizumi_TASLP_2023}), a text encoder (pre-trained RoBERTa \cite{Liu_arxiv_2019}), and a score predictor.
Given an audio spectrogram $\bm{X}$ and a text $\bm{w}$, the audio and text encoders produce the following representations:
\begin{align}
  \bmit{V} = \mathsf{AudioEnc}\left(\bm{X}\right) \in \mathbb{R}^{F \times T},\\
  \bm{o} = \mathsf{TextEnc}\left(\bm{w}\right) \in \mathbb{R}^{D}.
\end{align}
Then, $\bm{o}$ is duplicated in the time direction and concatenated with $\bmit{V}$.
The concatenated feature $\bmit{Z}\in \mathbb{R}^{F+D \times T}$ is then fed into the score predictor to predict the semantic alignment score $\hat{y}$.
\begin{align}
  \hat{y} = \mathsf{ScorePredictor}\left(\bmit{Z}\right),
\end{align}
The score predictor consists of a single-layer bidirectional long short-term memory (BiLSTM), a two linear layers, and the activation function ReLU.

During model training, the loss function $\mathcal{L}_{\mathrm{score}}$ was defined as a weighted linear combination of the clipped mean squared error (MSE) loss $\mathcal{L}_{\mathrm{mse}}$ and contrastive loss $\mathcal{L}_{\mathrm{con}}$ \cite{interspeech2025_kanamori_01}:
\begin{align}
  \mathcal{L}_{\mathrm{score}} = \mathcal{L}_{\mathrm{mse}} + \lambda_1 \mathcal{L}_{\mathrm{con}}
\end{align}

\subsection{Adversarial Training with a Gradient Reversal Layer}
The analysis in Section~\ref{sec:pre_exp} suggests that generated audio contains generation-model-specific characteristics that pretrained audio encoders can capture.
Such characteristics may hinder the generalization of semantic-alignment score prediction models to unseen audio generation models.
To address this issue, we introduce a GRL into the audio--text semantic-alignment score prediction model to encourage the learning of generation-model-invariant representations.

The GRL acts as an identity function during the forward pass but reverses and scales the gradients during backpropagation.
The scaling coefficient is annealed according to the DANN schedule~\cite{Ganin_JMLR_2016}.
A generation-model classifier consisting of a two-layer multilayer perceptron (MLP) is connected to an intermediate representation through the GRL.
The classifier is trained to identify the generating model using a cross-entropy loss.
The total training loss is defined as
\begin{equation}
\mathcal{L}_{\mathrm{total}} = \mathcal{L}_{\mathrm{score}} + \lambda_2 \mathcal{L}_{\mathrm{adv}},
\label{eq:total_loss}
\end{equation}
where $\mathcal{L}_{\mathrm{adv}}$ is the cross-entropy loss for generation-model classification and $\lambda_2$ is its weight.
The GRL reverses the gradient of $\mathcal{L}_{\mathrm{adv}}$ before it reaches the audio encoder, encouraging the encoder to learn representations that are useful for semantic-alignment score prediction but less informative about the generating model.

\section{Experiments}
\subsection{Experimental conditions}
\textbf{Dataset}.
We used the XACLE Challenge 2026 dataset \cite{Okamoto_ICASSP_2026} in our experiments.
The dataset consists of audio clips, text descriptions, and 11-point subjective audio--text alignment scores.
The audio includes both real-world recordings and audio generated by eight text-to-audio generation models.

To evaluate generalization to unseen audio generation models, we divided the eight models into two disjoint sets.
Set A consists of AudioLDM \cite{Liu_ICML_2023}, Tango2 \cite{Majumder_ACMMM_2024}, AudioGen \cite{Kreuk_ICLR_2023}, and Stable Audio\footnote{\url{https://stableaudio.com/}}, whereas Set B consists of AudioLDM2 \cite{Liu_TASLP_2024}, Tango \cite{Ghosal_ACMMM_2023}, Make-an-Audio 2 \cite{Huang_arXiv_2023}, and TangoFlux \cite{Hung_arXiv_2024}.
Models were trained on one set and evaluated on the other.

Each set contains 2,000, 1,000, and 1,000 audio--text pairs for training, validation, and test, respectively.
Because the number of samples varies across generation models, we randomly selected audio--text pairs from models with larger numbers of samples to construct a balanced dataset.

\textbf{Compared models}.
We compared the baseline model with a model incorporating a GRL.
For the GRL-based model, we evaluated two configurations: applying the GRL to the audio encoder (Only audio encoder) and applying it to the score predictor (All).

\textbf{Training settings.}
Each score prediction model was trained using the Adam optimizer with a learning rate of $2 \times 10^{-5}$, a batch size of 12, and 20 epochs.
The loss weights $\lambda_1$ and $\lambda_2$ were set to 0.5 and 0.1, respectively.
The GRL scaling coefficient $\alpha$ was fixed at 1.0 throughout training, while the adversarial loss weight $\lambda_2$ followed a two-stage schedule: 1.0 for the first 5 epochs, then reduced to 0.1 for the remaining epochs.
The model with the best validation SRCC was selected for evaluation.

\textbf{Evaluation metrics}.
We evaluated the proposed method using four evaluation metrics.
Spearman’s rank correlation coefficient (SRCC), Kendall’s rank correlation coefficient (KTAU), and the linear correlation coefficient (LCC) were used to measure the correlation between the predicted alignment scores and the subjective scores.
Mean squared error (MSE) was used to evaluate the prediction error between the predicted and subjective scores.

In addition to the four metrics described above, we evaluate how much generation-model-specific information is retained in the learned audio representations.
Following the protocol in Section~\ref{sec:pre_exp}, we train a two-layer MLP classifier using frozen audio-encoder representations from the baseline and GRL-based models.
The classifier is trained separately for Set A and Set B, each containing audio generated by four models.

\subsection{Results}
\textbf{Score prediction performance.}
Table~1 shows the semantic-alignment score prediction results for seen and unseen generation models.
When trained on Set A, the baseline achieved an SRCC of 0.339 on Set A but only 0.144 on the unseen Set B.
Similarly, when trained on Set B, the SRCC decreased from 0.393 on Set B to 0.210 on the unseen Set A.
These results indicate that the baseline model has limited generalization ability to unseen TTA models.

Applying the GRL improved prediction performance in some settings.
When trained on Set A, Ours (All) improved the SRCC from 0.339 to 0.413 on Set A and from 0.144 to 0.161 on the unseen Set B.
When trained on Set B, Ours (All) also improved the SRCC on the unseen Set A from 0.210 to 0.241.

The effect of the GRL application differed depending on the training set.
When trained on Set A, Ours (All) achieved the highest SRCC on both seen and unseen generation models.
In contrast, when trained on Set B, Ours (Only audio encoder) achieved the highest SRCC of 0.412 on Set B, while its performance on the unseen Set A was comparable to or slightly lower than that of the baseline.
These results suggest that the effectiveness of the GRL depends on its application location and the training data.

\textbf{Analysis of generation-model-specific information.}
To examine whether the GRL reduces generation-model-specific information in the learned audio representations, we evaluated generation-model classification accuracy using the classifier.
As shown in Table~\ref{tab:probe_accuracy}, Ours (All) reduced the classification accuracy from 0.942 to 0.835 when trained on Set B, whereas the accuracy increased from 0.928 to 0.951 when trained on Set A.
In contrast, Ours (Only audio enc.) did not reduce the classification accuracy for either set.
These results suggest that the effectiveness of the GRL in suppressing generation-model-specific information depends on the training data and its application location.

\begin{table}[t]
  \centering
  \footnotesize
  \setlength{\tabcolsep}{8pt}
  \caption{Audio generation model classification accuracy (4-class,
    matching Set A / Set B), using frozen audio-encoder
    representations before and after adversarial training.}
  \label{tab:probe_accuracy}
  \begin{tabular}{@{}clr@{}}
    \hline
    Training data & Model & Accuracy \\
    \hline
    \multirow{3}{*}{A} & Baseline   & 0.928 \\
     & Ours (All) & 0.951 \\
     & Ours (Only audio enc.) & 0.929 \\
    \hline
    \multirow{3}{*}{B} & Baseline   & 0.942 \\
     & Ours (All) & 0.835 \\
     & Ours (Only audio enc.) & 0.952 \\
    \hline
  \end{tabular}
\end{table}
\section{Conclusion}
In this study, we investigated the generalization of audio--text semantic-alignment score prediction models to unseen TTA models.
We first showed that pretrained audio encoders capture generation-model-specific characteristics and that score prediction performance degrades on audio generated by unseen TTA models.
To improve generalization, we introduced a GRL to encourage the learning of generation-model-invariant representations.
Experimental results showed that applying the GRL improved prediction performance in some settings, although the improvements on unseen TTA models were limited.
Furthermore, generation-model classification accuracy decreased only in one setting, indicating that the effectiveness of the GRL in suppressing generation-model-specific information was inconsistent.
Future work will explore more effective methods for reducing generation-model-specific information and improving generalization to unseen TTA models.


\begin{thebibliography}{99} 
\bibitem{Liu_TASLP_2024} H. Liu, Y. Yuan, X. Liu, X. Mei, Q. Kong, Q. Tian, Y. Wang, W. Wang, Y. Wang, and M. D. Plumbley, ``AudioLDM 2: Learning holistic audio generation with self-supervised pretraining,'' IEEE/ACM Transactions on Audio, Speech, and Language Processing, Vol. 32, pp. 2871--2883, 2024.

\bibitem{Lloyd_SIGG_2011} D. B. Lloyd, N. Raghuvanshi, and N. K. Govindaraju, ``Sound synthesis for impact sounds in video games,'' Proc. Symposium on Interactive 3D Graphics and Games, pp. 55--61, 2024.

\bibitem{Kong_ICASSP_2019} Q. Kong, Y. Xu, T. Iqbal, Y. Cao, W. Wang, and M. D. Plumbley, ``Acoustic scene generation with conditional sampleRNN,'' Proc. IEEE International Conference on Acoustics, Speech and Signal Processing (ICASSP), pp. 925--929, 2019.

\bibitem{Lee_NeurIPS_01} J. Lee and M. Tailleur and M. Lagrange and  K. Choi ane  L. Heller, B. McFee and K. Imoto and Y. Okamoto, ``Challenge on sound scene synthesis: Evaluating text-to-audio generation,'' Proc. Audio Imagination: NeurIPS 2024 Workshop AI-Driven Speech, Music, and Sound Generation, 2024.

\bibitem{interspeech2025_kanamori_01} Y. Kanamori and Y. Okamoto and T. Takano and S. Takamichi and Y. Saito and H. Saruwatari, ``RELATE: Subjective evaluation dataset for automatic evaluation of relevance between text and audio,'' Proc. INTERSPEECH, pp. 3155--3159, 2024.

\bibitem{Wu_ICASSP_2023} Y. Wu, K. Chen, T. Zhang, Y. Hui, T. Berg-Kirkpatrick, and S. Dubnov, ``Large-scale contrastive language-audio pretraining with feature fusion and keyword-to-caption augmentation,'' Proc. ICASSP, 2023.

\bibitem{Okamoto_ICASSP_2026} Y. Okamoto, R. Takizawa, M. Kishi, Y. Kanamori, N. Tonami, R. Nagase, S. Takamichi, and K. Imoto, ``XACLE Challenge 2026: The first x-to-audio alignment challenge,'' Proc. ICASSP, pp. 21877--21879, 2026.

\bibitem{Cooper_ICASSP_2022} E. Cooper, W. -C. Huang, T. Toda and J. Yamagishi, ``Generalization qbility of MOS prediction networks,'' Proc. ICASSP, pp. 8442--8446, 2022.

\bibitem{Huang_ICME_2026} K.-T. Huang, C.-C. Wang, C.-Y. Yang, H.-S. Lee, H.-M. Wang, B. Chen, ``Robust generative audio quality assessment: Disentangling quality from spurious correlations,'' Proc. IEEE International Conference on Multimedia and Expo, 2026.

\bibitem{Ganin_ICML_2015} Y. Ganin, V. Lempitsky, ``Unsupervised domain adaptation by backpropagation,'' Proc. International Conference on Machine Learning (ICML), pp. 1180--1189, 2015.

\bibitem{Niizumi_TASLP_2023} D. Niizumi, D. Takeuchi, Y. Ohishi, N. Harada, and K. Kashino, ``Byol for audio: Exploring pre-trained general purpose audio representations,'' IEEE/ACM Transactions on Audio, Speech, and Language Processing, vol. 31, pp. 137--151, 2023.

\bibitem{Liu_arxiv_2019} Y. Liu, M. Ott, N. Goyal, J. Du, M. Joshi, D. Chen, O. Levy, M. Lewis, L. Zettlemoyer, and V. Stoyanov, ``Roberta: A ro-
bustly optimized BERT pretraining approach,'' arXiv preprint arXiv:1907.11692, 2019

\bibitem{Ganin_JMLR_2016} Y. Ganin, E. Ustinova, H. Ajakan, P. Germain, H. Larochelle, F. Laviolette, M. Marchand, V. Lempitsky, ``Domain-adversarial training of neural networks,'' The Journal of Machine Learning Research, vol. 17, no. 1, pp. 2096--2030, 2016.

\bibitem{Liu_ICML_2023} H. Liu, Z. Chen, Y. Yuan, X. Mei, X. Liu, D. Mandic, W. Wang, and M. D. Plumbley, ``AudioLDM: Text-to-audio generation with latent diffusion models,'' Proc. ICML, pp. 21450--21474, 2023.

\bibitem{Majumder_ACMMM_2024} N. Majumder, C.-Y. Hung, D. Ghosal, W.-N. Hsu, R. Mihalcea, and S. Poria, ``Tango 2: Aligning diffusion-based text-to-audio generations through direct preference optimization,'' Proc. ACM International Conference on Multimedia (ACM MM), pp. 564--572, 2024.

\bibitem{Kreuk_ICLR_2023} F. Kreuk, G. Synnaeve, A. Polyak, U. Singer, A. D{\'e}fossez, J. Copet, D. Parikh, Y. Taigman, and Y. Adi, ``AudioGen: Textually guided audio generation,'' Proc. International Conference on Learning Representations (ICLR), 2023.

\bibitem{Ghosal_ACMMM_2023} D. Ghosal, N. Majumder, A. Mehrish, and S. Poria, ``Text-to-audio generation using instruction guided latent diffusion model,'' Proc. ACM International Conference on Multimedia (ACM MM), pp. 3590--3598, 2023.

\bibitem{Huang_arXiv_2023} J. Huang, Y. Ren, R. Huang, D. Yang, Z. Ye, C. Zhang, J. Liu, X. Yin, Z. Ma, and Z. Zhao, ``Make-an-Audio 2: Temporal-enhanced text-to-audio generation,'' arXiv preprint arXiv:2305.18474, 2023.

\bibitem{Hung_arXiv_2024}
C.-Y. Hung, N. Majumder, Z. Kong, A. Mehrish, A. Zadeh, C. Li, R. Valle, B. Catanzaro, and S. Poria,
``TangoFlux: Super fast and faithful text to audio generation with flow matching and CLAP-ranked preference optimization,''
arXiv preprint arXiv:2412.21037, 2024.

\end{thebibliography}

\end{document}